\documentclass[a4paper,fleqn]{article}  
 \usepackage{amsfonts}
 \usepackage{latexsym}
 \usepackage{doc}
 \usepackage{fontenc}
 \usepackage{exscale}
 \usepackage{graphics}
 \newcommand{\mb}{\mbox}
 \newcommand{\nw}{\newcommand}
 \nw{\hs}{\hspace}
 \nw{\vs}{\vspace}
 
\begin{document}
 % \begin{titlepage}
 \title{New classes of quasi-solvable potentials,their exactly-solvable limit % 
                            and related orthogonal polynomials}                
 \author{Asish Ganguly\thanks{e-mail:asish@cucc.ernet.in} \\
 Department of Applied Mathmatics, University of Calcutta \\
 92 Acharya Prafulla Chandra Road, Kolkata --- 700009, India}
 \date{3 May 2002 \\ Published in Nov.\ JMP issue}
 \maketitle
 \begin{abstract}
 We have generated, using an sl(2,$\mathbb{R}$) Lie-algebraic formalism several new classes of 
 quasi-solvable elliptic potentials, which in the appropriate limit go over to the exactly solvable forms. 
 We have obtained exact solutions of the corresponding spectral problem for some  real values 
 of the potential  parameters. We have also given  explicit expressions of the families of associated 
 orthogonal polynomials in the energy variable.
 \end{abstract}
 % \end{titlepage}
 %P.S. \hspace{.1cm} Running title of the paper could be ``QES potentials and related orthogonal 
  %            \\ \hspace*{1cm} polynomials''
 %\vspace{1.5cm}
 %\pagebreak
 \setcounter{equation}{0}
 \renewcommand{\theequation}{1. \arabic{equation}}
 \section{Introduction}

 In recent times elliptic potentials have proved to be an important addition [1--3] to 
 the class of solvable \cite{lev,bb:book} and quasi-solvable [6--8] potentials in quantum 
 mechanics. In particular, within the sl(2,$\mathbb{R}$) algebra,exact solutions of Lam\'{e} and associated 
 Lam\'{e} equation have been obtained [9--12] for various ranges of the potential parameters. 
 Indeed a handful of theorems relating to the properties of elliptic potentials
 are known for a long time [13--16] including the study of the properties of the corresponding 
 wavefunctions \cite{mag}. The solutions of associated Lam\'{e} equation have also been obtained \cite{kh} 
 by using these theorems which, however, do not use the sl(2,$\mathbb{R}$) technique. Note that some new 
 elliptic models based on Weierstrass $\wp$ function have recently been 
 proposed \cite{shif3} wherein it is shown that the corresponding Hamiltonians  possess the so-called 
 energy-reflection symmetry \cite{dunne3}. 

 By the term quasi-exactly solvable(QES) periodic potentials we mean potentials consisting of finite number 
 of allowed bands and expressible as  doubly-periodic elliptic functions which are either Jacobian 
 elliptic functions $snx\equiv sn(x,k), cnx\equiv cn(x,k), dnx \equiv dn(x,k)$ of real elliptic 
 modulus parameter $k(0<k^{2}<1)$ or Weierstrass $\wp$ function. This is in sharp 
 contrast with the ordinary ES periodic potentials with a single period. 

 There is an intriguing relation between ES and QES class. In fact an sl(2) based construction with an
 $n$-dimensional finite space representation gives ($n$+1) levels for a Hamiltonian designated as QES
 model. It was pointed out in Ref.\ 8 that if one can construct a Hamiltonian having no explicit dependence
 on $n$, then in the limit $n\rightarrow \infty$ the ES models are recovered. This provides a sufficient
 reason to believe that corresponding to every ES model there ought to exist a QES model which in 
 the proper limit goes over to the former.

 In this article we show that the elliptic parameter $k$ can be used for passage from QES to ES in the 
 periodic models. We derive three new QES periodic potentials involving Jacobian elliptic functions.
 The elliptic functions having real and imaginary period reduce to ordinary periodic functions namely
 hyperbolic(imaginary period) and trigonometric(real period) functions as the modulus parameter $k$
 goes to 1 and 0. We exploit this interesting property of elliptic functions to show that our QES models
 are connected with some well known ES periodic class. Note that we do not intend to take the limit
 $n\rightarrow \infty$ and as such we cannot expect to recover the whole spectrum for ES models. 
 Rather we show that the limit $k\rightarrow 1$ and $0$ correctly map the few lower states of the QES
 and ES periodic potentials.

 The plan of this article is as follows. In Sec.\ 2 we briefly review the basics of the sl(2,$\mathbb{R}$) 
 Lie-algebraic formalism and generate Type I,Type II and Type III models within this framework. The method of
 construction of the related orthogonal polynomials is also sketched here. Specific examples are constructed 
 in Sec.~3 for each of them based on some real values of the potential parameters. In Sec.~4 we 
 systematically analyze the ES-limit of our results to show how this limit can reproduce the ES results. 
 Finally we present our conclusions in Sec.\ 5.
 \setcounter{equation}{0}
 \renewcommand{\theequation}{2. \arabic{equation}}
 \section{New QES potentials from sl(2,$\mathbb{R}$) and related orthogonal polynomials}

 To start with, let us adopt the following differential realization of the sl(2,$\mathbb{R}$) generators
 \begin{equation}
 \hspace{2cm} T^{+}=\xi^{2}\partial_{\xi} - n\xi, \quad T^{0}= \xi\partial_{\xi} -\frac{1}{2}n, \quad 
                                           T^{-}=\partial_{\xi},
 \end{equation}
 obeying commutation relations
 \begin{equation}
 \hspace{2cm}[T^{+},T^{-}] = -2T^{0}, \qquad [T^{0},T^{\pm}] = \pm T^{\pm},
 \end{equation}
 where $n$ is a non-negative integer. The gauged Hamiltonian is taken as the standard homogeneous quadratic 
 combination of sl(2,$\mathbb{R}$) generators along with linear terms:
 \begin{equation}
 H_{G}=-C_{++}T^{+^{2}}-C_{00}T^{0^{2}}-C_{--}T^{-^{2}}-C_{+}T^{+}-C_{0}T^{0}-C_{-}T^{-}-d,
 \end{equation}
 where $C_{ii}$,$C_{j}$($i,j=0,\pm$) are numerical parameters and $d$ is a suitably chosen constant 
 taken as function of $C_{j}$. Note that $d$ plays the role of an  overall shift in the energy scale. This 
 psudo degree of freedom allows us to obtain QES models in the desired form.

 Substitution of (2.1) into (2.3) yields
 \begin{eqnarray}
 H_{G}(\xi) & = & -(C_{++}\xi^{4}+C_{00}\xi^{2}+C_{--})\partial_{\xi}^{2}  
 -[2(1-n)C_{++}\xi^{3} +C_{+}\xi^{2}+\{(1-n)C_{00} \nonumber \\
            &   & \mbox{}+ C_{0} \}\xi+C_{-}] \, \partial_{\xi}  
             -\! [n(n-1)C_{++}\xi^{2}-\! nC_{+}\xi \! +\frac{n^{2}}{4}C_{00} \! -\frac{n}{2}C_{0} \! +d \, ], 
 \end{eqnarray}
 which after a coordinate transformation
 \begin{equation}
 \hspace{4cm}x(\xi) = \int^{\xi}d\tau/ \sqrt{C_{++}\tau^{4}+C_{00}\tau^{2}+C_{--}} \, ,
 \end{equation}
 converts $H_{G}$ into the form
 \begin{eqnarray}
 H_{G}(x) & = & -\partial^{2}_{x}+ \frac{2nC_{++}\xi^{3}(x)-C_{+}\xi^{2}(x)+(nC_{00}-C_{0})\xi(x)-C_{-}}
       {\sqrt{C_{++}\xi^{4}(x)+C_{00}\xi^{2}(x)+C_{--}}} \, \partial_{x} \nonumber \\
          &   & \mbox{} -[n(n-1)C_{++}\xi^{2}(x)-nC_{+}\xi(x)+\frac{n^{2}}{4}C_{00}-\frac{n}{2}C_{0}+d \, ],
 \end{eqnarray}
 where $\xi=\xi(x)$ is determined by (2.5).

 Let us now consider the Schr\"{o}dinger equation with the potential $V(x)$
 \begin{equation}
 \hspace{2cm}H(x)\psi(x) \equiv [-\partial_{x}^{2}+V(x)]\psi(x) = E\psi(x)
 \end{equation}
 Writing $\psi(x)$ in the form
 \begin{equation}
 \hspace{4cm} \psi(x) = \mu(x)\chi(x) ,
 \end{equation}
 we obtain
 \begin{equation}
 H_{G}(x)\chi(x) \equiv [-\partial^{2}_{x}-2(\frac{\mu^{'}}{\mu})\partial_{x} - (\frac{\mu^{'}}{\mu})^{2}
              - (\frac{\mu^{'}}{\mu})^{'} + V ] \, \chi(x) = E\chi(x).
 \end{equation} 

 Comparing (2.6) and (2.9) we find for the potential $V(x)$ and the gauge factor $\mu(x)$ the following 
 relationships
 \begin{equation}
 V(x)=(\frac{\mu^{'}}{\mu})^{2}+(\frac{\mu^{'}}{\mu})^{'}-[n(n-1)C_{++}\xi^{2}-nC_{+}\xi
                       +\frac{n^{2}}{4}C_{00}-\frac{n}{2}C_{0}+d \, ] ,
 \end{equation}
 \begin{equation}
 \mu(x)= \! [C_{++}\xi^{4}+C_{00}\xi^{2}+C_{--}]^{-\frac{n}{4}}\exp \, [\int^{\xi} \! \frac{C_{+}\tau^{2}
              +C_{0}\tau+C_{-}}{2(C_{++}\tau^{4}+C_{00}\tau^{2}+C_{--})}d\tau] .
 \end{equation}

 Note that the choice of numerical parameters $C_{ii}$ must be such that equation (2.5) may be invertible 
 in terms of \( \xi=\xi(x)\). For our purpose $\xi(x)$ needs to be expressed in terms of 
 Jacobian elliptic functions. In Ref.~12 we gave an almost exhaustive list of the choice of $C_{ii}$ leading 
 to various new classes of elliptic potentials. Here we consider the following three types of combinations of 
 parameters namely
 \begin{eqnarray}
 \mbox{Type I :} \qquad C_{++}=-k^{2}      \quad & C_{00}=2k^{2}-1    \quad & C_{--}=k^{'^{2}} , \\
 \mbox{Type II:} \qquad C_{++}=k^{2}      \quad & C_{00}=-(1+k^{2})  \quad & C_{--}=1 , \\
 \mbox{Type III}: \qquad C_{++}=k^{'^{2}} \quad & C_{00}=1+k^{'^{2}} \quad & C_{--}=1 ,
 \end{eqnarray}
 where $k^{2}\in(0,1)$ and $k^{'^{2}}=1-k^{2}$. Each of the above types defines different coordinate
 transformations through (2.5). These give respectively $\xi=-cnx,\,-cnx/dnx$ and $snx/cnx$
 for the three types mentioned above.

 In this way we are then led to the following new classes of elliptic potentials:
 \begin{eqnarray}
 \hspace{-1cm} \mbox{Type I :}  & V(x)= & [B^{2}+A(A+1)]\frac{dn^{2}x}{sn^{2}x}
                                    -2B(A+\frac{1}{2})\frac{cnx}{sn^{2}x}, \quad x\in(0,2K) \\
 \hspace{-1cm} \mbox{Type II:}  & V(x)= & B(B+1)\frac{dn^{2}x}{sn^{2}x}-A(A+1)dn^{2}x, 
                                                \quad \qquad x\in(0,2K) \\
 \hspace{-1cm} \mbox{Type III}: & V(x) \! = & \hspace{-.2cm} [B^{2} \! - \! A(A+1)]k^{2}cn^{2}x+ \! 2Bk^{2}
                                                          (A+\frac{1}{2})snxcnx, \, x\in(-\infty,\infty)
 \end{eqnarray}
 where $K = \int^{\pi/2}_{0}d\alpha/\sqrt{1-k^{2}\sin^{2}\alpha}$ is the complete elliptic integral 
 of 1st kind. Note that in (2.15)--(2.17) the potential parameters $A,B\in \mathbb{R}$ and the choices of 
 $C_{j}$ and the spin parameter $n$ in (2.10)  in terms of $A,B$ are given in Table 1. 
 
 \nw{\hl}{\hs{.9cm}}
 \begin{table}[t]
 \small
 \caption{\small{Different algebraizations for Type I--III potentials are given. Last column gives
    restrictions on potential parameters to keep $n$ to a non-negative integer}}

 \vs{1mm}
 \noindent
 \begin{tabular}{@{}cc@{\hs{-.1pt}}lllll@{}} \hline \hline
 Type & Solution &    &  & & & Restrictions \\ 
name & \hs{.2cm}no. & \hs{.4cm}$n$ & \hl$C_{+}$ & \hl$C_{-}$ & \hs{.5cm}$C_{0}$ & \hs{.1cm}on $A,B$ \\ \hline
 I & 1.1 & $A$ & $2k^{2}B$ & $2k^{'^{2}}B$ & $A$ & $A \! \in \! \mathbb{N}-1,B \! \in \! \mathbb{R}$ \\
& 1.2 & $A \! -1$ & $2k^{2}B$ & $2k^{'^{2}}B$ & $A \! +1$ & $A \! \in \! \mathbb{N},B \! \in \! \mathbb{R}$ \\
        & 1.3      & $B \! -1$ & $2k^{2}(A \! +\frac{1}{2}) \! -ikk^{'}$      
& $2k^{'^{2}}(A \! +\frac{1}{2}) \! +ikk^{'}$  & $B$ & $A \! \in \! \mathbb{R},B \! \in \! \mathbb{N}$ \\
& 1.4 & $A \! -\frac{1}{2}$ & $2k^{2}B \! -ikk^{'}$ 
& $2k^{'^{2}}B \! +ikk^{'}$ & $A \! +\frac{1}{2}$ & $A \! \in \! \mathbb{N} \! -\frac{1}{2},
                                             B \! \in \! \mathbb{R}$ \\ 
& 1.5 & $B \! -\frac{1}{2}$   & $2k^{2}(A \! +\frac{1}{2})$             
& $2k^{'^{2}}(A \! +\frac{1}{2})$ & $B \! -\frac{1}{2}$ & $A \! \in \! \mathbb{R},
                                           B \! \in \! \mathbb{N} \! -\frac{1}{2}$ \\
& 1.6      & $B \! -\frac{3}{2}$ & $2k^{2}(A \! +\frac{1}{2})$ & $2k^{'^{2}}(A \! +\frac{1}{2})$ & 
 $B \! +\frac{1}{2}$ & $A \! \in \! \mathbb{R},B \! \in \! \mathbb{N} \! +\frac{1}{2}$ \\
II & 2.1 & $A \! -\frac{1}{2}$ & $- \! 2k^{2}(B \! +\frac{1}{2})$ & $2(B \! +\frac{1}{2})$ & 
 $- \! k^{'^{2}} \! (A \! +\frac{1}{2})$ & $A \! \in \! \mathbb{N} \! -\frac{1}{2},B \! \in \! \mathbb{R}$ \\
& 2.2 & $A \! -\frac{1}{2}$ & $2k^{2}(B \! +\frac{1}{2})$ & $- \! 2(B \! +\frac{1}{2})$ & 
 $- \! k^{'^{2}} \! (A \! +\frac{1}{2})$ & $A \! \in \! \mathbb{N} \! -\frac{1}{2},B \! \in \! \mathbb{R}$ \\
III & 3.1 & $A$ & $- \! 2k^{'^{2}}B$ & $- \! 2B$ & $- \! Ak^{2}$ & $A \! \in \! \mathbb{N}-1,
                                        B \! \in \! \mathbb{R}$ \\  
& 3.2 & $A \! -1$ & $- \! 2k^{'^{2}}B$ & $- \! 2B$ & $- \! (A \! +1)k^{2}$ & 
                                         $A \! \in \! \mathbb{N},B \! \in \! \mathbb{R}$ \\
& 3.3 & $A \! -\frac{1}{2}$ & $- \! 2k^{'^{2}}B \! +ik^{'}$ & $- \! 2B \! +ik^{'}$  & 
                       $- \! (A \! +\frac{1}{2})k^{2}$ & 
                         $A \! \in \! \mathbb{N} \! -\frac{1}{2},B \! \in \! \mathbb{R}$ \\ \hline \hline
 \end{tabular}
 \end{table}

 %################################################################
 The constant d is chosen as
 \begin{eqnarray}
 \mbox{Type I}:   & d= & \frac{C_{0}}{4}\{C_{0}-4k^{'^{2}}(n+1)\}+\frac{C_{+}}{2k^{2}}(k^{2}C_{-}
                                    -k^{'^{2}}C_{+})+\frac{n(n+2)}{2}k^{'^{2}} \\
   &   &                                                 \nonumber \\
 \mbox{Type II}:  & d= & (\frac{C_{+}}{2k})^{2}-\frac{C_{0}}{2}(n+1)+\frac{n(n+2)}{4}(1+k^{2}) \\
   &   &                                                  \nonumber \\
 \mbox{Type III}: & d= & \frac{1}{4}[(\frac{C_{+}}{k^{'}})^{2}-n(n+2)(2-k^{2})-2C_{0}(n+1)] 
 \end{eqnarray}  

 Note that while the Type III potential is defined over the entire $x$-axis, Type I and Type II potentials 
 are singular at $x=0,2K$  and so defined over an open domain $(0,2K)$. It follows from 
 the oscillation theorem that we need to find the periodic solutions [of period $4K$(or $8K$)] for 
 $4K$-periodic potential of Type I and periodic solutions [of period $2K$(or $4K$)] for $2K$-periodic 
 potentials of Type II and Type III at $E=E_{j}$. The monotonic increasing sequence $\{E_{j}\}$, where 
 $E_{0} < E_{1}\leq E_{2} < E_{3}\leq E_{4} \cdots ,$ gives the characteristic values of the 
 energy parameter.

 Following the  analysis made in \cite{as2} we can find the effective combinations of the potential
 parameters $A,B$. We see that the Type I and Type III potentials are invariant under the translation
 $A,B\rightarrow A^{'},B^{'}\) where \(A^{'}=-A-1,B^{'}=-B$ and Type II potential is invariant
 under $A,B\rightarrow A^{'},B^{'}$ where $A^{'}=-A-1,B^{'}=-B-1$. Further due to the periodic relations
 $sn(x+2K)=-snx, \, cn(x+2K)=-cnx, \, dn(x+2K)=dnx,$ it results that the effective regions in the $A$-$B$ 
 plane for Type I--III models are bounded by the constraints $A\geq-1/2,B\geq 0 ;A,B\geq -1/2$ and 
 $A\geq -1/2,B\in \mathbb{R}$ respectively. It may be pointed out that the eigenstates and spectra of Type I 
 potential for $B<0$ can be obtained from those for $B>0$ under the coordinate translation 
 $x \rightarrow x+2K$.

 Before concluding this section let us briefly describe the method of construction of families of orthogonal 
 polynomials \cite{as2,ben,fin2} generated by the eigenstates of a QES Hamiltonian. Let us consider
 a gauged eigenvalue equation
 \begin{equation}
 \hspace{4cm} H_{G}(\xi)\chi(\xi) = E\chi(\xi)  
 \end{equation}
 where we identify \( \chi(\xi(x)) \equiv \chi(x) \) as given in (2.8). We now expand the gauged 
 eigenfunction $\chi(\xi)$ in the form
 \begin{equation}
 \hspace{4cm} \chi(\xi)=(\frac{\xi_{2}-\xi}{\xi_{1}-\xi_{2}})^{n}\sum_{j=0}^{\infty}\frac{P_{j}(E)}{j!}
(                                                             \frac{\xi_{1}-\xi}{\xi_{2}-\xi})^{j}
 \end{equation}
 where $\xi_{1},\xi_{2}$ are two distinct roots of the coeffecient of $\partial_{\xi}^{2}$ in (2.4).
 Now a suitable choice of $\xi_{1},\xi_{2}$ (note that $\xi_{1},\xi_{2}$ can be chosen in six ways) 
 gives us a  three-term recursion relation satisfied by 
 \{$P_{j}(E)$\}
 \begin{eqnarray}
 -[(2j-n+1)\hat{C}_{0-}+\hat{C}_{-}]P_{j+1} & \! = & \! [E+d_{1}+\hat{C}_{0}(j-\frac{n}{2})
                      +\hat{C}_{00}(j-\frac{n}{2})^{2}] \!P_{j} \nonumber \\
                                            &      & \mbox{}+j(j-1-n)[ \! (2j-n-1)\hat{C}_{+0}
                                                        +\hat{C}_{+}] \! P_{j-1} ,  \nonumber \\
                                            &      & \hspace{2.1in} (j\geq 0),
 \end{eqnarray}
 where $\hat{C}_{ij}$ are determined from the relations
 \begin{eqnarray}
 \hat{C}_{+0} & = & -\frac{1}{(\xi_{1}-\xi_{2})^{2}}[2\xi_{1}\xi_{2}^{3}C_{++}+\xi_{2}(\xi_{1}+\xi_{2})
                    C_{00}+2C_{--}], \nonumber \\
 \hat{C}_{00} & = & \frac{1}{(\xi_{1}-\xi_{2})^{2}}[6\xi_{1}^{2}\xi_{2}^{2}C_{++}+(\xi_{1}^{2}
                    +\xi_{2}^{2}+4\xi_{1}\xi_{2})C_{00}+6C_{--}], \nonumber \\
 \hat{C}_{0-} & = & -\frac{1}{(\xi_{1}-\xi_{2})^{2}}[2\xi_{1}^{3}\xi_{2}C_{++}+\xi_{1}(\xi_{1}+\xi_{2})
                    C_{00}+2C_{--}], 
 \end{eqnarray}
 and $\hat{C}_{j},d_{1}$ are given by
 \begin{eqnarray}
 \hat{C}_{+}  & = & \frac{1}{\xi_{1}-\xi_{2}}[\xi_{2}^{2}C_{+}+\xi_{2}C_{0}+C_{-}], \nonumber \\
 \hat{C}_{0}  & = & -\frac{1}{\xi_{1}-\xi_{2}}[2\xi_{1}\xi_{2}C_{+}+(\xi_{1}+\xi_{2})C_{0}
                                                        +2C_{-}], \nonumber \\
 \hat{C}_{-}  & = & \frac{1}{\xi_{1}-\xi_{2}}[\xi_{1}^{2}C_{+}+\xi_{1}C_{0}+C_{-}], \nonumber \\
 d_{1}        & = & d+\frac{n(n+2)}{12}(C_{00}-\hat{C_{00}}).
 \end{eqnarray}

 From equation (2.23)--(2.25) it transpires that the eigenstates of Type~I--III Hamiltonians generate,in 
 general, different orthogonal family of polynomials in the energy variable corresponding to each
 algebraization of Table 1 provided 
 \begin{equation}
 \hspace{4cm} (2j-n+1)\hat{C}_{0-}+\hat{C}_{-}\neq 0,\qquad \forall \, j\geq 0
 \end{equation}

 The family \{$P_{j}$(E)\} can be expressed in terms of monic polynomials \{$\tilde{P}_{j}$(E)\}
 satisfying the recurrence relation
 \begin{eqnarray}
 \hspace{4cm} \tilde{P}_{j+1} & = & (E-\lambda_{j})\tilde{P}_{j}-\rho_{j}\tilde{P}_{j-1} \, , \\
 \hspace{4cm} \tilde{P}_{j}   & = & \omega_{j}P_{j} \, , \qquad j\geq 0, 
 \end{eqnarray}
 where $\tilde{P}_{-1}=P_{1}\equiv 0$ and $\tilde{P}_{0}=P_{0}\equiv 1$. It is now straightforward
 to write down the expression of eigenfunctions from (2.8) and (2.22). It follows from equation (2.23) 
 that $\rho_{0}=0$ and $\rho_{n+1}=0$; so the infinite power series expansion in (2.22) truncates
 after the ($n$+1)-th term since the coeffecients $P_{j}(E_{i})$ vanishes for $j > n$, where 
 $E_{i} \, (i=0,1, \ldots ,n)$ are the zeros of the critical polynomial $\tilde{P}_{n+1}(E)$. This points to 
 the fact that Type~I--Type~III potentials belong to QES class having $(n+1)$ levels for each
 algebraization. The final expression of the band-edge eigenfunctions may be written in the form
 \begin{equation}
 \psi_{E_{i}}(x)=\mu(x)(\xi(x)-\xi_{2})^{n}\sum_{j=0}^{n}\frac{P_{j}(E_{i})}{j!}
                             (\frac{\xi(x)-\xi_{1}}{\xi(x)-\xi_{2}})^{j}, \quad (i=0,1, \ldots ,n),
 \end{equation}
 where $\mu(x)$ is determined from (2.11) for each of the three types given by (2.12)--(2.14) and the 
 band-edge eigenvalues are $E_{0},E_{1}, \cdots E_{n}$. Note that $n$ is to be computed for each of the 
 algebraization in Table 1.
 \setcounter{equation}{0}
 \renewcommand{\theequation}{3. \arabic{equation}}
 \section{Eigenstates and spectra of Type I--III models for some real values of the potential parameters}

 In this section we construct some examples based upon the general results obtained in the previous
 section. It may be useful to collect the following identities and differential relations among 
 the Jacobian elliptic functions which will be frequently used:
 \begin{displaymath}
 \hspace{4cm} sn^{2}x+cn^{2}x=1, \: dn^{2}x+k^{2}sn^{2}x=1
 \end{displaymath}
 \begin{displaymath}
 \hs{2cm}sn^{'}x=cnxdnx,\, cn^{'}x=-snxdnx, \, dn^{'}x=-k^{2}snxcnx
 \end{displaymath}        

 In the following examples we denote the eigenstates[spectra] by $\phi_{r}(x)[e_{r}]$ whenever 
 ordering is possible. Otherwise we denote them by $\psi_{E_{i}}[E_{i}]$ and $\psi_{E_{i}^{'}}[E^{'}_{i}]$
 indicating different algebraizations. 
 \subsection{Type I model [defined on the domain (0,2K)]}

 We have got six algebraizations (see the solution 1.1--1.6 of Table 1) for Type I Hamiltonian (2.15). For
 each of them the corresponding eigenstates generate an orthogonal family of polynomials satisfying the 
 recurrence relation (2.27). The explicit expressions of $\rho_{j}$ and $\lambda_{j}$ corresponding to
 each solutions of $n$ are given in Table 2. The corresponding expressions of $\omega_{j}$ together with 
 the choice of $\xi_{1},\xi_{2}$ and the overall restrictions on potential parameters are given in Table 3. 
 \begin{table}[t]
  \small
 \caption{The coeffecients $\rho_{j}$ and $\lambda_{j}$ of the recurrence relation (2.27) are
 given for each of the six algebraizations of Type I model.}

 \vs{1mm}
 \noindent
 \begin{tabular}{@{}clll@{}} \hline \hline
 Solution & & & \\
 \hs{1mm}no. & \hs{.1cm} $n$   & \hs{2cm} $\rho_{j}$ & \hs{2.3cm} $\lambda_{j}$ \\ \hline
 1.1 & $A$ & $(\frac{1}{2})^{2}j(j-1-A)(2j+2B-1)$ & $\frac{1-2k^{2}}{2}[A(A+1)+(A-2j)(2B-A+2j)]$ \\
  &  & \hs{.2cm}$\times (2B-2A+2j-1)$ & \\ 
 1.2 & $A-1$ & $(\frac{1}{2})^{2}j(j-A)(2j+2B+1)$ & $\frac{1-2k^{2}}{2}[A(A+1)+(A-1-2j)$ \\
  &  & \hs{.2cm}$\times (2B-2A+2j-1)$ & \hs{1cm}$\mbox{}\times (2B+2j-A+1)]$ \\ 
 1.3  & $B-1$ & $(\frac{1}{2})^{2}j(j-B)(2j+1)$ & $\frac{1-2k^{2}}{4}[2B^{2}-1+2(B-1-2j)(2j-B+2)]$ \\ 
 & & \hs{.2cm}$\times(2j-2B+1)$ & \hs{.2cm}$\mbox{}+\frac{1}{2}ikk^{'}(2A+1)(2B-2A-4j-3)$ \\ 
 1.4 & $A-\frac{1}{2}$ & $(\frac{1}{2})^{2}j(2j-2A-1)(j-A)$ & $\frac{1-2k^{2}}{8}
  [4A(A+1)-1+(2A-4j-1)$ \\ 
  & & \hs{.2cm}$\mbox{}\times (2j+1)$  & \hs{.2cm}$\mbox{}\times (4j-2A+3)]+2Bikk^{'}(A-2j-1)$ \\  
 1.5  & $B-\frac{1}{2}$ & $(\frac{1}{2})^{2}j(j+A)(2j-2B-1)$ & $\frac{1-2k^{2}}{8}
  [4B^{2}-1+(2B-4j-1)$ \\
  & & \hs{.2cm}$\times (2A-2B+2j+1)$ & \hs{1cm}$\mbox{}\times (4j+4A-2B+3)]$ \\ 
 1.6 & $B-\frac{3}{2}$ & $(\frac{1}{2})^{2}j(2j-2B+1)(j+A+1)$ & $\frac{1-2k^{2}}{8}
  [4B^{2}-1+(2B-4j-3)$ \\ 
  & & \hs{.2cm}$\times (2A-2B+2j+1)$ & \hs{1cm}\mbox{}$\times (4j+4A-2B+5)]$ \\ \hline \hline
 \end{tabular}
 \end{table}

%############################################################

 It is clear that the algebraic solutions are obtained for the following two cases:
 \begin{displaymath}
 \hspace{-1cm} \mbox{Case 1.} \qquad A\in (\mathbb{N}-1)\cup (\mathbb{N}-\frac{1}{2}), B\in \mathbb{R}
 \end{displaymath}

 Here $B$ is any real parameter and for each real values of $B$, $A$ is allowed to take 
 $0,1/2,1,3/2, \cdots ,\Lambda \leq B$($\Lambda$ is an integer or half-integer). For integer values of 
 $A$, from the algebraizations 1.1--1.2 we get $(2A+1)$ band edge eigenstates and eigenvalues. Also 
 for half-integer values of $A$ the algebraization 1.4 gives $(A+1/2)$ solutions of the Schr\"{o}dinger 
 equation. 
 \begin{displaymath}
 \hspace{-1cm} \mbox{Case 2.} \qquad B\in \mathbb{N}\cup \mathbb{N}-\frac{1}{2}, A\in \mathbb{R}
 \end{displaymath}

 Here $A$ is any real number and $B$ is allowed to take values $1/2,1,3/2,\cdots ,\Lambda \leq A+1$($\Lambda$
 is integer or half-integer). It is to be noted that algebraization 1.3 is considered for integer values 
 of $B$ and 1.5--1.6 for half-integer $B$.
  \begin{table}[t]
 \small
 \caption{The expressions for $\omega_{j}$ of the relation (2.28) along with the choice of roots 
 $\xi_{1},\xi_{2}$ and the overall restrictions on potential parameters are provided for each
 algebraization for Type I model.}

 \vs{1mm}
 \noindent
 \renewcommand{\arraystretch}{1.6}
 \begin{tabular}{@{}cllll@{}} \hline \hline
 Solution & & & & Overall restrictions \\
 \hs{1mm}no. & $n$ & \hs{1.3cm} $\omega_{j}$ & $(\xi_{1},\xi_{2})$ & \hs{.1cm} on $A,B$   \\ \hline
 1.1 & $A$ &
 $(\frac{1}{2})^{j}\frac{\prod_{r=0}^{[B]-A+j}(2B-2A+2j-1-2r)}{\prod_{r=0}^{[B]-A}(2B-2A-1-2r)}$ &
         $(-1,1)$ & $A\in \mathbb{N}-1,B\in \mathbb{R} \, ,B-A\geq 0$ \\
 1.2 & $A-1$ &
 $(\frac{1}{2})^{j}\frac{\prod_{r=0}^{[B]-A+j}(2B-2A+2j-1-2r)}{\prod_{r=0}^{[B]-A}(2B-2A-1-2r)}$ &
        $(-1,1)$ & $A\in \mathbb{N},B\in \mathbb{R} \, ,B- A\geq 0$ \\
 1.3 & $B-1$ &
 $(\frac{-1}{2})^{j}\prod_{r=0}^{j}(2j+1-2r)$ & $(\frac{ik'}{k},\frac{-ik'}{k})$ & 
                                 $A\in \mathbb{R},B\in \mathbb{N} \, ,A- B+1\geq 0$ \\
 1.4 & $A-\frac{1}{2}$ &
 $(\frac{-1}{2})^{j}\prod_{r=0}^{j}(2j+1-2r)$ & $(\frac{ik'}{k},\frac{-ik'}{k})$ & 
                $A\in \mathbb{N}-\frac{1}{2},B\in \mathbb{R} \, ,B- A\geq 0$ \\
 1.5 & $B-\frac{1}{2}$ & 
 $(\frac{1}{2})^{j}\frac{\prod_{r=0}^{[A-B]+j}(2A-2B+2j-2r+1)}{\prod_{r=0}^{[A-B]}(2A-2B-2r+1)}$ & $(-1,1)$
                & $A\in \mathbb{R},B\in \mathbb{N}-\frac{1}{2} \, ,A- B+1\geq 0$ \\
 1.6 & $B-\frac{3}{2}$ &
 $(\frac{1}{2})^{j}\frac{\prod_{r=0}^{[A-B]+j}(2A-2B+2j-2r+1)}{\prod_{r=0}^{[A-B]}(2A-2B-2r+1)}$ & $(-1,1)$
  & $A\in \mathbb{R},B\in \mathbb{N}+\frac{1}{2} \, ,A- B+1\geq 0$ \\ \hline \hline
 \end{tabular}
 \renewcommand{\arraystretch}{1}
 \end{table}

 %################################################

 We now consider some specific examples when both of $A$ and $B$ are integer or half-integer.
 \begin{displaymath}
 \hspace{-1cm} \mbox{{\bf 1. A = 0,B=1/2 :}} \qquad V(x) \, = \, \frac{1}{4}\frac{dn^{2}x}{sn^{2}x} 
                                                              -\frac{1}{2}\frac{cnx}{sn^{2}x}
 \end{displaymath}
 \begin{equation}
 \phi_{0}(x) \, = \, \sqrt{\frac{snx}{1+cnx}}, \qquad e_{0} \, = \, 0
 \end{equation}
 \begin{displaymath}
 \hspace{-1cm} \mbox{{\bf 2. A=0,B=3/2 :}} \qquad V(x) \, =  \, \frac{9}{4}\frac{dn^{2}x}{sn^{2}x} 
                                                                  -\frac{3}{2}\frac{cnx}{sn^{2}x}
 \end{displaymath}
 \begin{equation}
 \phi_{0}(x) \, = \, (\frac{snx}{1+cnx})^{3/2}, \qquad e_{0} \, = \, 0
 \end{equation}
 \begin{displaymath}
 \hspace{-1cm} \mbox{{\bf 3. A = 1/2,B=1 :}} \qquad V(x) \, = \, \frac{7}{4}\frac{dn^{2}x}{sn^{2}x}
                                                    -2\frac{cnx}{sn^{2}x}
 \end{displaymath}
 \begin{equation}
 \phi_{0}(x) \, = \, \frac{\sqrt{snxdnx}}{1+cnx}exp \, (-\frac{i}{2}tan^{-1}\frac{kcnx}{k^{'}}), \qquad
                                          e_{0} \, = \, \frac{1-2k^{2}}{4}-ikk^{'}
 \end{equation}
 \begin{displaymath}
 \hspace{-1cm} \mbox{{\bf 4. A=1/2,B=2 :}} \qquad V(x) \, = \, \frac{19}{4}\frac{dn^{2}x}{sn^{2}x}
                                                           -4\frac{cnx}{sn^{2}x}
 \end{displaymath}
 \begin{equation}
 \phi_{0}(x) \, = \, \frac{\sqrt{dnx}sn^{3/2}x}{(1+cnx)^{2}}exp \, (-\frac{i}{2}tan^{-1}\frac{kcnx}{k^{'}}),
                            \qquad e_{0} \, = \, \frac{1-2k^{2}}{4}-2ikk^{'}
 \end{equation}
 \begin{displaymath}
 \hspace{-1cm} \mbox{{\bf 5. A = 1,B=3/2 :}} \qquad V(x) \, = \, \frac{17}{4} \, \frac{dn^{2}x}{sn^{2}x} 
                                               -\frac{9}{2}\frac{cnx}{sn^{2}x}
 \end{displaymath}
 For $0 \, < \, k^{2} \, < \, 1/2 \, $,
 \begin{equation}
 \hspace{-1cm}\phi_{0,1}(x)=\frac{\sqrt{snx}}{(1+cnx)^{5/2}}[g_{\mp}(k)cn^{2}x+4cnx+4-g_{\mp}(k)], 
                       \: e_{0,1}=1-4k^{2}+\frac{1}{2}g_{\mp}(k),
 \end{equation}
 \begin{equation}
 \hspace{1cm}\phi_{2}(x)=\frac{dnx\sqrt{snx}}{(1+cnx)^{3/2}}, \qquad e_{2}=1-2k^{2},                         
 \end{equation}
 \begin{displaymath}
 \mbox{where} \qquad g_{\pm}(k) \, = \, 6k^{2}-1\pm\sqrt{1-36k^{2}k^{'^{2}}}
 \end{displaymath}
 For $1/2 \, < \, k^{2} \, < \, 1 \, $,the suffixes 0,1,2 have to be replaced by 1,2,0 respectively.

 Proceeding in the same fashion,we can find the eigenstates and spectra for higher values of A and B.
 \subsection{Type II model [defined on the domain (0,2K)]}

 Here two algebraizations are obtained (see the solution 2.1--2.2 of Table 1). The related orthogonal
 polynomials are determined by the entries 2.1--2.2 of Table 4 and Table 5. %
  \begin{table}[t]
  \small
 \caption{The coefficients $\rho_{j}$ and $\lambda_{j}$ of the recurrence relation (2.27) are provided
 for Type II(1st two rows) and Type III models.}
 
 \vs{1mm}
 \noindent
 \begin{tabular}{@{}clll@{}} \hline \hline
 Solution & & & \\
  \hs{.2cm}no. & \hs{.1cm} $n$ & \hs{2cm} $\rho_{j}$ & \hs{2.3cm} $\lambda_{j}$ \\ \hline
 2.1 & $A-\frac{1}{2}$ & $(\frac{k^{'^{2}}}{2})^{2}j(2j-2A-1)(j-A+B)$ & $-\frac{k^{2}}{4}(2B+1)^{2}
                                                 -\frac{k^{'^{2}}}{8}(2A+1)^{2}+\frac{1+k^{2}}{8}$ \\ 
  & & \hs{.2cm}$\times (2j+2B+1)$ & \hs{.2cm}$\times (2A-4j-1)(4B-2A+4j+3)$  \\
 2.2 & $A-\frac{1}{2}$ & $(\frac{k^{'^{2}}}{2})^{2}j(2j-2A-1)(j-A-B-1)$ & $-\frac{k^{2}}{4}(2B+1)^{2}
                                                      -\frac{k^{'^{2}}}{8}(2A+1)^{2}+\frac{1+k^{2}}{8}$ \\ 
  & & \hs{.2cm}$\times (2j-2B-1)$ & \hs{.2cm}$\times(2A-4j-1)(4j-2A-4B-1)$ \\ 
   & & & \\  
 3.1 & $A$ & $(\frac{k^{2}}{2})^{2}j(j \! -A \! -1)(2j \! -1)(2j \! -2A \! -1)$ & $\frac{2j-A}{2}
     [(2j-A)(2-k^{2})+4Bik^{'}]$ \\
 & & & \hs{.2cm}$-B^{2}k^{'^{2}}-A(A+1)k^{2}/2$  \\
 3.2 & $A-1$ & $(\frac{k^{2}}{2})^{2}j(j \! -A)(2j \! -2A \! -1)(2j \! +1)$ & $\frac{2j-A+1}{2}
                                                                   [(2j-A+1)(2-k^{2})+4Bik^{'}]$ \\ 
 & & & \hs{.2cm}$-B^{2}k^{'^{2}}-A(A+1)k^{2}/2$  \\
 3.3 & $A-\frac{1}{2}$ & $(\frac{k^{2}}{2})^{2}j(j-A)(2j-2A-1)(2j+1)$ & $(\frac{1+2Bik^{'}}{2})^{2}
                                        -\frac{k^{2}}{8}(2A \! +1)^{2}+\frac{1}{8}(2A \! -4j \! -1)$ \\ 
 & & & \hs{.2cm}$\times[(k^{2}-2)(4j-2A+3)-8Bik^{'}]$  \\ \hline \hline
 \end{tabular}
 \end{table}

%####################################################
 Note that both algebraizations are valid provided $A$ is restricted to positive half-integer values only 
while the other parameter $B$ is  arbitrary.

 Some examples are furnished below.
 \begin{equation}
 {\bf 1. \: A \, = \, \frac{1}{2}}: \qquad V(x) \, = \, B(B+1)\frac{dn^{2}x}{sn^{2}x}-\frac{3}{4}dn^{2}x
 \end{equation}
 \begin{displaymath}
 \phi_{0}(x) \, = \, \frac{sn^{B+1}x}{(cnx+dnx)^{B+1/2}}, \qquad e_{0} \, = \, 
                           -\frac{k^{'^{2}}}{2}-\frac{k^{2}}{4}(2B+1)^{2}
 \end{displaymath}
 For $B \, \leq \, 0 \, $, the following degenerated state is found:
 \begin{displaymath}
 \psi_{0}(x) \, = \, (cnx+dnx)^{B+1/2}/sn^{B}x
 \end{displaymath}
 \begin{equation}
 {\bf 2. \: A \, = \, \frac{3}{2}}: \qquad V(x) \, = \, B(B+1)\frac{dn^{2}x}{sn^{2}x}-\frac{15}{4}dn^{2}x
 \end{equation}
 \begin{eqnarray*}
 \phi_{0,1}(x) & = & \frac{sn^{B+1}x}{(cnx+dnx)^{B+3/2}}[\{(1+k^{2})(B+\frac{3}{2})+\eta_{\mp}(k)\}cn^{2}x \\ 
               &   & \mbox{}+2(B+\frac{3}{2})cnxdnx+k^{'^{2}}(B+\frac{3}{2})-\eta_{\mp}(k)],  
 \end{eqnarray*}
 \begin{displaymath}
 e_{0,1}=\frac{6k^{2}-10-k^{2}(2B+1)^{2}}{4}\mp \sqrt{2(1+k^{4})(B+\frac{1}{2})^{2}-k^{'^{4}}},
 \end{displaymath}
 \begin{displaymath}
 \mbox{where} \qquad \eta_{\pm}(k) \, = \, -(1+k^{2})(B+\frac{1}{2})\pm \sqrt{2(1+k^{4})(B+\frac{1}{2})^{2}
                                                                  -k^{'^{4}}}
 \end{displaymath}
 For $B \, \leq \, 0 \, $ two other degenerated states are obtained:
 \begin{eqnarray*}
 \psi_{0,1}(x) & = & \frac{(cnx+dnx)^{B-1/2}}{sn^{B}x}[\{(1+k^{2})(B+\frac{3}{2})+\eta_{\mp}(k)\}cn^{2}x \\
               &   & \mbox{}+(1-2B)cnxdnx-k^{'^{2}}(B-\frac{1}{2})-(1+k^{2})(2B+1)-\eta_{\mp}(k)]
 \end{eqnarray*} 
 \subsection{Type III model [defined on the entire real line]}
 
 This corresponds to three algebraizations (see the solution 3.1--3.3 of Table 1), the first two of which 
 are for an integer $A$ while the third one is for an half-integer $A$. As before, the eigenstates generate 
 an orthogonal family of polynomials in the energy variable for each algebraization. %
  \begin{table}[t]
 \small
 \caption{The expressions for $\omega_{j}$ of the relation (2.28) together with the choice of 
 $\xi_{1},\xi_{2}$ and the overall restrictions on potential parameters are provided for Type II
 and Type III models.}

 \vs{1mm}
 \noindent
 \renewcommand{\arraystretch}{1.2}
 \addtolength{\tabcolsep}{2mm}
 \begin{tabular}{@{}cllll@{}} \hline \hline
 Solution & & & & Overall Restrictions \\
 \hs{.2cm}no. & \hs{1mm}$n$ & \hs{1.3cm}$\omega_{j}$ & $(\xi_{1},\xi_{2})$ & \hs{1mm} $A,B$ \\ \hline
 2.1 & $A-\frac{1}{2}$ & $(\frac{k'^{2}}{2})^{j}\frac{\prod_{r=0}^{j+[B]}(2j+2B-2r+1)}
   {\prod_{r=0}^{[B]}(2B-2r+1)}$ & $(-1,1)$ & $A\in \mathbb{N}-\frac{1}{2},B\in \mathbb{R}$ \\
 2.2 & $A-\frac{1}{2}$ & $(\frac{k'^{2}}{2})^{j}\frac{\prod^{j}_{r=0}(2j-2B-2r-1)}{-2B-1}$ 
         & $(-1,1)$ & $A\in \mathbb{N}-\frac{1}{2},B\in \mathbb{R},B\leq 0$ \\
 & & & & \\
 3.1 & $A$ & $(\frac{k^{2}}{2})^{j}\frac{\prod^{j}_{r=0}(2j-1-2r)}{(-1)}$ & 
  $(\frac{i}{k'},\frac{-i}{k'})$ & $A\in \mathbb{N}-1,B\in \mathbb{R}$ \\
 3.2 & $A-1$ & $(\frac{k^{2}}{2})^{j}\prod^{j}_{r=0}(2j+1-2r)$ & $(\frac{i}{k'},\frac{-i}{k'})$ &
                      $A\in \mathbb{N},B\in \mathbb{R}$ \\
 3.3 & $A-\frac{1}{2}$ & $(\frac{k^{2}}{2})^{j}\prod^{j}_{r=0}(2j+1-2r)$ & $(\frac{i}{k'},\frac{-i}{k'})$ &
                              $A\in \mathbb{N}-\frac{1}{2},B\in \mathbb{R}$ \\ \hline \hline
 \end{tabular}
 \renewcommand{\arraystretch}{1}
 \addtolength{\tabcolsep}{-2mm}
 \end{table}

 The recurrence relation (2.27) is determined by the entries 3.1--3.3 of Table 4 and Table 5. The other 
 parameter $B$ takes  arbitrary  values.

 We now consider some examples.
 \begin{equation}
 {\bf 1. \: A \, = \, 0}: \qquad V(x) \, = \, B^{2}k^{2}cn^{2}x+Bk^{2}snxcnx
 \end{equation}
 \begin{displaymath}
 \phi_{0}(x) \, = \, exp \, (-Btan^{-1}\frac{snx}{cnx}), \qquad e_{0} \, = \, -B^{2}k^{'^{2}}
 \end{displaymath}
 \begin{equation}
 {\bf 2. \: A \, = \, \frac{1}{2}}: \qquad V(x) \, = \, (B^{2}-\frac{3}{4})k^{2}cn^{2}x+2Bk^{2}snxcnx
 \end{equation}
 \begin{displaymath}
 \phi_{0}(x)=\sqrt{dnx}exp \, [-Btan^{-1}(\frac{snx}{cnx})+\frac{i}{2}tan^{-1}(\frac{k^{'}snx}{cnx})], \,
                                e_{0}=-\frac{k^{2}}{2}+(\frac{1+2Bik^{'}}{2})^{2}
 \end{displaymath}
 \begin{equation}
 {\bf 3. \: A \, = \, 1}: \qquad V(x) \, = \, (B^{2}-2)k^{2}cn^{2}x+3Bk^{2}snxcnx
 \end{equation}
 \begin{displaymath}
 \phi_{0}(x)=dnxexp \, (-Btan^{-1}\frac{snx}{cnx}), \qquad e_{0}=-B^{2}k^{'^{2}}-k^{2}
 \end{displaymath}
 \begin{displaymath}
 \phi_{1,2}^{(i)}(x)=[(k^{2}\mp \sqrt{k^{4}-16k^{'^{2}}B^{2}})snx+4Bcnx]
                        exp \, (-Btan^{-1}\frac{snx}{cnx}), 
 \end{displaymath}
 \begin{displaymath}
 \phi_{1,2}^{(ii)}(x)=[4Bk^{'^{2}}snx+(k^{2}\pm \sqrt{k^{4}-16k^{'^{2}}B^{2}})cnx]
                                 exp \, (-Btan^{-1}\frac{snx}{cnx}),
 \end{displaymath}
 \begin{displaymath}
 \hspace{3cm} e_{1,2}=1-\frac{3}{2}k^{2}-B^{2}k^{'^{2}}\mp \frac{1}{2}\sqrt{k^{4}-16k^{'^{2}}B^{2}}, 
 \end{displaymath}
 where the superscripts in the eigenstates indicate their double-degeneracy.
 \setcounter{equation}{0}
 \renewcommand{\theequation}{4. \arabic{equation}}
 \section{ES-limit of QES models}

 We have so far constructed three new classes of QES potentials and  explicitly obtained their 
 eigenstates and spectra. Our purpose in this section is to show that corresponding to each type there 
 is associated an ES class potential. It is  useful to write down the following results of the ES-limit
 \begin{eqnarray}
 sn(x,k)  \renewcommand{\arraystretch}{.3}
                 \begin{array}{l}
   \scriptstyle{k \! \rightarrow \! 1} \\ \longrightarrow \\ \scriptstyle{k \! \rightarrow \! 0}
                 \end{array} 
           \renewcommand{\arraystretch}{1}
                                 \left \{ \begin{array}{l}
                                         \tanh x \\ \sin x
                               \end{array} \right., &
 cn(x,k) \renewcommand{\arraystretch}{.3}
                 \begin{array}{l}
   \scriptstyle{k \! \rightarrow \! 1} \\ \longrightarrow \\ \scriptstyle{ k \! \rightarrow \! 0}
                  \end{array}
          \renewcommand{\arraystretch}{1}
                                 \left \{ \begin{array}{l}
                                     \mb{sech}x \\ \cos x
                                \end{array} \right., &
 dn(x,k)  \renewcommand{\arraystretch}{.3}
                  \begin{array}{l}
    \scriptstyle{k \! \rightarrow \! 1} \\ \longrightarrow \\ \scriptstyle{k \! \rightarrow \! 0} 
                  \end{array}
          \renewcommand{\arraystretch}{1}
                           \left \{ \begin{array}{l}
                                  \mb{sech}x \\ 1 
                                \end{array} \right.
 \end{eqnarray}
 \renewcommand{\arraystretch}{1}

 Each of the three types of potentials are doubly periodic, one is real and the other imaginary. As 
 $k\rightarrow 1$(or 0) we get an ES potential having an imaginary(or real) period.
 \subsection{ES classes associated to Type I model}

 We have already shown that the Type I model [cf.\ equation (2.15)] belongs to the QES periodic class. 
 The potential has a real period $4K$ and imaginary period $4K^{'}$, where $K^{'}\equiv K(k^{'})$. 
 Using the relation (4.1) and the relation
 \begin{eqnarray}
                     \hspace{2cm} &
 K \: [K^{'}] \renewcommand{\arraystretch}{.3}
                         \begin{array}{l}
 \scriptstyle{k \! \rightarrow \! 1} \\ \longrightarrow \\ \scriptstyle{k \! \rightarrow \! 0}
                          \end{array}
                \renewcommand{\arraystretch}{1}
                        \left \{ \begin{array}{l}
                             \infty \: [\pi/2] \\ \pi/2 \: [\infty]
                                 \end{array} \right. &
              \begin{array}{l}
                       \hspace{2cm} \\ ,
              \end{array}
 \end{eqnarray}
 we see that the QES model is also exactly solvable when the modulus parameter $k\rightarrow 1$ and 0.
 The associated ES classes are given by 
 \begin{equation}
 \hspace{-.5cm}k\rightarrow 1:V_{1}(x) \!= \! [B^{2}+A(A+1)]\mb{cosech}^{2}x
                 -2B(A+\frac{1}{2})\mb{cosech}x\coth x,  \,    x\in (0,\infty ) 
 \end{equation}
 \begin{equation}
 \hspace{-.5cm} k\rightarrow 0:V_{2}(x)= [B^{2}+A(A+1)]\mb{cosec}^{2}x-2B(A+\frac{1}{2})\mb{cosec}x\cot x, 
                                                                            \quad x\in (0,\pi ), 
 \end{equation}
 whose eigenstates and spectra are \cite{lev} 
 \begin{eqnarray}
 \psi_{r}^{(1)}(x) & = & (\cosh x-1)^{\frac{B-A}{2}}(\cosh x+1)^{-\frac{B+A}{2}}
        P_{r}^{(B-A-\frac{1}{2},-B-A-\frac{1}{2})}(\cosh x), \nonumber \\
 \psi_{r}^{(2)}(x) & = & (1-\cos x)^{\frac{B-A}{2}}(1+\cos x)^{-\frac{B+A}{2}}
        P_{r}^{(B-A-\frac{1}{2},-B-A-\frac{1}{2})}(\cos x),
 \end{eqnarray}
 \begin{equation}  
 E_{r}^{(1)}=-(A-r)^{2}, \qquad E_{r}^{(2)}=(A-r)^{2}, \qquad (r=0,1,2, \cdots )
 \end{equation}
 where $P_{r}^{(\alpha,\beta)}(x)$ is the Jacobi polynomial and the 
 superscripts 1 and 2 indicates the potentials $V_{1}(x)$ and $V_{2}(x)$ respectively. Note
 that we have solved the Type I spectral problem(see Subsec.\ 3.1) when at least one of the parameters 
 $A,B$ is an integer or an half-integer. Thus the ES results (4.5) and (4.6) can be reproduced from the 
 associated QES results for this restricted domain of potential parameters only.

 We now consider the ES-limits of the examples given in Subsec.\ 3.1 [see equations (3.1)--(3.5)]. In the 
 following the superscripts in the eigenstates indicate whether they correspond to the potentials 
 $V_{1}(x)$ and $V_{2}(x)$.     
 \begin{displaymath}
 {\bf 1. \: A \, = \, 0,B=1/2 :}
 \end{displaymath}
 \begin{eqnarray}   
  \hspace{2cm} V_{1}(x) & = & \frac{1}{4}\mb{cosech}^{2}x-\frac{1}{2}\mb{cosech}x\coth x \\
  \hspace{2cm} V_{2}(x) & = & \frac{1}{4}\mb{cosec}^{2}x-\frac{1}{2}\mb{cosec}x\cot x
 \end{eqnarray}
 \begin{displaymath}
 \phi_{0}^{(1)}(x)=\sqrt{\tanh\frac{x}{2}}, \qquad \phi_{0}^{(2)}(x)=\sqrt{\tan\frac{x}{2}}, \qquad 
                                   e_{0}^{(1)}=e_{0}^{(2)}=0
 \end{displaymath}
 
 From (4.5) and (4.6) it follows that we can reproduce the ground states for the two ES classes.
 \begin{displaymath}
 {\bf 2. \: A \, = \, 0,B=3/2 :}
 \end{displaymath} 
 \begin{eqnarray}
  \hspace{2cm} V_{1}(x) & = & \frac{9}{4}\mb{cosech}^{2}x-\frac{3}{2}\mb{cosech} x\coth x \\
  \hspace{2cm} V_{2}(x) & = & \frac{9}{4}\mb{cosec}^{2}x-\frac{3}{2}\mb{cosec}x\cot x
 \end{eqnarray}
 \begin{displaymath}
 \phi_{0}^{(1)}(x)=\tanh^{\frac{3}{2}}\frac{x}{2}, \qquad \phi_{0}^{(2)}(x)=\tan^{\frac{3}{2}}\frac{x}{2}, 
                                        \qquad e_{0}^{(1)}=e_{0}^{(2)}=0
 \end{displaymath}
 Again ground levels for two ES potentials (4.9),(4.10) are reproduced.
 \begin{displaymath}
 {\bf 3. \: A \, = \, 1/2,B=1 :}
 \end{displaymath}
 \begin{eqnarray}
 \hspace{2cm} V_{1}(x) & = & \frac{7}{4}\mb{cosech}^{2}x-2\mb{cosech} \, x\coth x \\
 \hspace{2cm} V_{2}(x) & = & \frac{7}{4}\mb{cosec}^{2}x-2\mb{cosec} \, x\cot x
 \end{eqnarray}
 \begin{displaymath}
 \phi_{0}^{(1)}(x)=\frac{\sqrt{\mb{sech} \, x\tanh x}}{1+\mb{sech} \, x}, \qquad 
       \phi_{0}^{(2)}(x)=\frac{\sqrt{\sin x}}{1+\cos x},    \qquad e_{0}^{(1)}=-\frac{1}{4}=-e_{0}^{(2)}
 \end{displaymath}
 These are the ground states for the ES classes.
 \begin{displaymath}
 {\bf 4. \: A \, = 1/2,B=2 :}
 \end{displaymath}
 \begin{eqnarray}
  \hspace{2cm} V_{1}(x) & = & \frac{19}{4}\mb{cosech}^{2}x-4\mb{cosech} \, x\coth x \\
  \hspace{2cm} V_{2}(x) & = & \frac{19}{4}\mb{cosec}^{2}x-4\mb{cosec} \, x\cot x
 \end{eqnarray}
 \begin{displaymath}
 \phi_{0}^{(1)}(x)=\frac{\sinh^{3/2}x}{(1+\cosh x)^{2}}, \qquad \phi_{0}^{(2)}(x)=
    \frac{\sin^{3/2}x}{(1+\cos x)^{2}}, \qquad e_{0}^{(1)}=-\frac{1}{4}=-e_{0}^{(2)}
 \end{displaymath}
 These are the ground levels for ES potentials.
 \begin{displaymath}
 {\bf 5. \: A \, = \, 1,B=3/2 :}
 \end{displaymath}
 \begin{eqnarray}
  \hspace{2cm} V_{1}(x) & = & \frac{17}{4}\mb{cosech}^{2}x-\frac{9}{2}\mb{cosech} \, x\coth x \\
  \hspace{2cm} V_{2}(x) & = & \frac{17}{4}cosec^{2}x-\frac{9}{2}\mb{cosec} \, x\cot x
 \end{eqnarray}
 \begin{displaymath}
 \phi_{0}^{(1)}(x)=\frac{\sqrt{\sinh x}}{(1+\cosh x)^{3/2}}, \, \phi_{1}^{(2)}(x)= 
                 \frac{\sqrt{\sin x}}{(1+\cos x)^{3/2}}, \quad e_{0}^{(1)}=-1=-e_{1}^{(2)} 
 \end{displaymath}
 \begin{displaymath}
 \phi_{1}^{(1)}(x)=\frac{\sqrt{\sinh x}(\cosh x-3)}{(1+\cosh x)^{3/2}}, \, \phi_{0}^{(2)}(x)=
                          \frac{\sqrt{\sin x}(\cos x-3)}{(1+\cos x)^{3/2}},\quad  e_{1}^{(1)}=0=e_{0}^{(2)}
 \end{displaymath}
 Here two ES levels are reproduced.
 \subsection{ES class associated to Type II model}

 Using as before the relations (4.1) and (4.2), we see that in the limit $k\rightarrow 1$ Type II potential
 (2.16) goes over to the following ES class known as the generalized P\"{o}schl-Teller potential:
 \begin{equation}
 \hspace{1cm} V_{3}(x) \: = \: B(B+1)\mb{cosech}^{2}x-A(A+1)\mb{sech}^{2}x , 
 \end{equation}
 its eigenstates and spectra being given by
 \begin{displaymath}
 \psi_{r}^{(3)}(x) \: = \: (\cosh 2x-1)^{-B/2}(\cosh 2x+1)^{-A/2}
                                P_{r}^{(-B-\frac{1}{2},-A-\frac{1}{2})}(\cosh 2x),
 \end{displaymath}
 \begin{equation}
 E_{r}^{(3)}=-(A+B-2r)^{2}, \qquad (r = 0,1,2, \cdots ),
 \end{equation}
 where the superscripts indicate correspondance with $V_{3}(x)$.

 We recall that the QES levels for Type II model is obtained for half-integer values of $A$. Let us now 
 take the ES-limit of the examples given in Subsec.\ 3.2 [see equations (3.6)--(3.7)]. The other parameter
  $B$ is taken as negative real number.
 \begin{equation}
 {\bf 1. \, A=\frac{1}{2}:}  \qquad V_{3}(x)= B(B+1)\mb{cosech}^{2}x-\frac{3}{4}\mb{sech}^{2}x 
 \end{equation}
 \begin{displaymath}
 \psi_{0}^{(3)}(x)=\mb{cosech}^{B}x\sqrt{\mb{sech} \, x}, \qquad e_{0}^{(3)}=-\frac{(2B+1)^{2}}{4}.
 \end{displaymath}
 Here $\psi_{0}^{(3)},e_{0}^{(3)}$ are ground level for the ES potential $V_{3}(x)$.
 \begin{equation}
 {\bf 2. \, A=\frac{3}{2}:} \qquad V_{3}(x)=B(B+1)\mb{cosech}^{2}x-\frac{15}{4}\mb{sech}^{2}x
 \end{equation}
 \begin{displaymath}
 \psi_{0}^{(3)}(x)=\sinh^{-B} \! x \, \cosh^{-3/2}x, \qquad e_{0}^{(3)}=-(B+\frac{3}{2})^{2},
 \end{displaymath}
 \begin{displaymath}
\psi_{1}^{(3)}(x)=\frac{\sinh^{-B}x}{\cosh^{3/2}x}[(2B+1)\cosh^{2}x-2] \qquad 
                                               e_{1}^{(3)}=-(B-\frac{1}{2})^{2}
 \end{displaymath}
 Here $\psi_{0,1}^{(3)},e_{0,1}^{(3)}$ are two levels for the ES potential (4.20).
 \subsection{ES class associated to Type III model}
 
 The Type III potential (2.17) is defined on the entire real line. This is a QES periodic potential 
 having a real period $2K$ and an imaginary period $4K^{'}$. The algebraic sector is determined for 
 an integer or a half-integer $A$, while $B$ is arbitrary real parameter. In the limit $k\rightarrow 1$, 
 this potential coincides to the following ES class: 
 \begin{equation}
 \hspace{1cm} V_{4}(x) \, = \, [B^{2}-A(A+1)]\mb{sech}^{2}x+2B(A+\frac{1}{2})\mb{sech} \, x \, \tanh x.
 \end{equation}
 The eigenstates and spectra of the potential (4.21) are 
 \begin{displaymath}
 \psi_{r}^{(4)}(x)=\mb{sech}^{A}x\exp [-B\tan^{-1}(\sinh x)]P_{r}^{(-iB-A-\frac{1}{2},iB-A-\frac{1}{2})}
                                                        (i\sinh x),
 \end{displaymath}
 \begin{equation}
 E_{r}^{(4)}=-(A-r)^{2}, \qquad (r=0,1,2, \cdots )
 \end{equation}

 We now take  the ES-limit (here $k\rightarrow 1$ only) of the examples given in Subsec 3.3. Note that
 the limit $k\rightarrow 0$ gives a free particle motion.
 \begin{equation}
 {\bf 1. \, A=0:} \qquad V_{4}(x)=B^{2}\mb{sech}^{2}x+B\mb{sech} \, x \, \tanh x
 \end{equation}
 \begin{displaymath}
 \hspace{2cm} \phi_{0}^{(4)}(x)=\exp(-B\tan^{-1}\sinh x), \qquad e_{0}^{(4)}=0
 \end{displaymath}
 Clearly, the ground state for (4.23) agrees with the general results(4.22) for $A=0$.
 \begin{equation}
 {\bf 2. \, A=\frac{1}{2}:} \qquad V_{4}(x)=(B^{2}-\frac{3}{4})\mb{sech}^{2}x+2B\mb{sech} \, x \, \tanh x
 \end{equation}
 \begin{displaymath}
 \hspace{2cm} \phi_{0}^{(4)}(x)=\sqrt{\mb{sech} \, x}\exp[-B\tan^{-1}(\sinh x)], \qquad e_{0}^{(4)}=-1/4
 \end{displaymath}
 This is also in agreement with (4.22) for $A=1/2$.
 \begin{equation}
 {\bf 3. \, A=1:} \qquad V_{4}(x)=(B^{2}-2)\mb{sech}^{2}x+3B\mb{sech} \, x \, \tanh x
 \end{equation}
 \begin{eqnarray*}
 \phi_{0}^{(4)}(x)=\mb{sech} \, x\exp(-B\tan^{-1}\sinh x)=\phi_{1}^{(4)}(x) & , & e_{0}^{(4)}=-1=e_{1}^{(4)} 
 \end{eqnarray*}
 \begin{displaymath}
 \phi_{2}^{(4)}(x)=\mb{sech} \, x(\sinh x+2B)\exp(-B\tan^{-1}\sinh x), \qquad e_{2}^{(4)}=0
 \end{displaymath}
 Thus two ES levels are reproduced from the corresponding QES levels.
 
 Hence we have shown that corresponding to three QES models Type I,Type II and Type III, there is associated 
 some definite ES classes namely $V_{1},V_{2}$ (for TypeI),$V_{3}$(for Type II) and $V_{4}$(for Type III)
 respectively. In the ES limit we can reproduce some ES levels from the corresponding QES levels as well.
 \section{Conclusion}
 
 To conclude, we have constructed three new QES elliptic potentials Type I--III using 
 sl(2,$\mathbb{R}$) Lie-algebraic scheme and obtained their algebraic levels analytically. Further we have 
 shown that the eigenstates of QES Hamiltonians generate an orthogonal family of polynomials in the energy 
 variable. The interesting point is that the elliptic parameter k($0<k^{2}<1$) in the models turns out to be
 responsible for QES class and when it touches the end-points 0 and 1 of its domain the ES clsses are 
 revealed. We have explicitly shown that some ES levels can be reproduced on a restricted domain of 
 potential parameters.
 \section*{Acknowledgement}

 I would like to thank Professor B Bagchi for guidance.     
     
 \end{document}